\documentclass[sigconf]{acmart}

\renewcommand\footnotetextcopyrightpermission[1]{} 
\usepackage{color,soul}

\AtBeginDocument{%
  \providecommand\BibTeX{{%
    \normalfont B\kern-0.5em{\scshape i\kern-0.25em b}\kern-0.8em\TeX}}}

\begin{document}

\title{SliceHub: Augmenting Shared 3D Model Repositories with Slicing Results for 3D Printing}


\author{Faraz Faruqi}
\affiliation{%
  \institution{MIT CSAIL}
  \city{Cambridge, MA}
  \country{USA}}
\email{ffaruqi@mit.edu}

\author{Kenneth Friedman}
\affiliation{%
  \institution{MIT CSAIL}
  \city{Cambridge, MA}
  \country{USA}}
\email{ksf@mit.edu}

\author{Leon Cheng}
\affiliation{%
  \institution{MIT CSAIL}
  \city{Cambridge, MA}
  \country{USA}}
\email{leonc@mit.edu}

\author{Michael Wessely}
\affiliation{%
  \institution{MIT CSAIL}
  \city{Cambridge, MA}
  \country{USA}}
  \email{wessely@mit.edu}

\author{Sriram Subramanian}
\affiliation{%
  \institution{University College London}
  \city{London}
  \country{United Kingdom}}
  \email{s.subramanian@ucl.ac.uk}

\author{Stefanie Mueller}
\affiliation{%
  \institution{MIT CSAIL}
  \city{Cambridge, MA}
  \country{USA}}
  \email{stefanie.mueller@mit.edu}


\begin{abstract}
In this paper, we explore how to augment shared 3D model repositories, such as \textit{Thingiverse}, with slicing results that are readily available to all users. By having print time and material consumption for different print resolution profiles and model scales available in real-time, users are able to explore different slicing configurations efficiently to find the one that best fits their time and material constraints. To prototype this idea, we build a system called SliceHub, which consists of three components: (1) a repository with an evolving database of 3D models, for which we store the print time and material consumption for various print resolution profiles and model scales, (2) a user interface integrated into an existing slicer that allows users to explore the slicing information from the 3D~models, and (3)~a computational infrastructure to quickly generate new slicing results, either through parallel slicing of multiple print resolution profiles and model scales or through interpolation. We motivate our work with a formative study of the challenges faced by users of existing slicers and provide a technical evaluation of the SliceHub system. 
\end{abstract}



\begin{teaserfigure}
  \includegraphics[width=\textwidth]{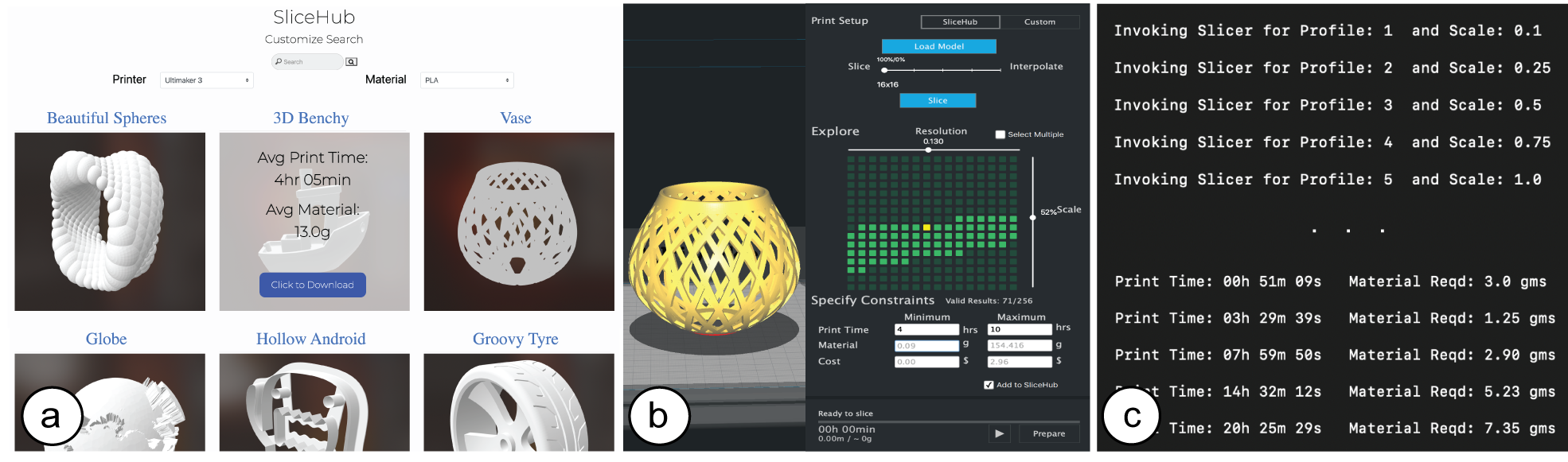}
  \caption{SliceHub’s integrated system: (a)~repository with slicing results, (b)~user interface for exploring trade-offs between different print resolution profiles and model scales, (c)~infrastructure for slicing and interpolation to generate new slicing results—the results can then be added to the repository further extending the available options.}
  \label{fig:1-main-figure}
\end{teaserfigure}

\maketitle

\section{Introduction}

3D printing a digital model typically requires a user to make certain choices: users have to select the 3D~printer and filament they would like to use, choose a print resolution profile, and scale the model to the desired size. This is done in a \textit{slicer}, a program that uses the selected print settings to convert the 3D model into a set of machine instructions that the 3D printer can execute.

The required print time and material consumption are strongly influenced by these print settings. For instance, an object printed with a higher print resolution profile requires more layers and thus more print time than the same object printed with a lower print resolution profile that requires fewer layers. Similarly, the scale of the model is an important factor for print time and material consumption since larger models require more layers and more material per layer. Since 3D printing is slow and can take many hours even for hand-sized objects, users often find themselves in a situation where they have to trade-off between different print resolution profiles and model scales~\cite{ludwig2014towards, hudson2016understanding}. 

Unfortunately, today, information on the required print time and material consumption for different print resolution profiles and model scales is not readily available to users. Since slicers first have to slice the model before they can display the expected print time and material consumption, users have to wait after each change of the print resolution profile and model scale. Since a single slicing process can take up to several minutes for complex 3D models, it makes the exploration of suitable print resolution profiles and model scales a time-consuming process. 

In this paper, we present an integrated system that enables users to explore different print settings in real-time to find the trade-off between print resolution profile, model scale, print time and material consumption that best fits their needs. Our idea is to augment 3D model repositories, such as \textit{Thingiverse}, where the same 3D model is downloaded and printed by hundreds of users~\cite{makerbot}, with slicing results that have to be computed only once and subsequently are stored and made available to all users of the repository. This allows users to explore multiple print settings simultaneously in real-time without having to wait for slicing to finish.

While our long-term vision is to integrate the slicing results into existing 3D model platforms, such as \textit{Thingiverse,} these commercial platforms are hard to extend with functionality due to limitations with their APIs. We therefore built the SliceHub repository to be able to prototype and study a system that stores and reuses slicing results. The repository contains for each 3D model, the print time and material consumption for different print resolution profiles and model scales, which users can download through each 3D model's page. 

While current slicers restrict users to explore print resolution profile and model scales one at a time, having the print time and material consumption for different settings available in real-time allows users to explore multiple options simultaneously. To support this exploration, we developed a user interface embedded within an existing slicer that provides users with an overview of the print time and material consumption resulting from different print resolution profiles and model scales. The user interface also supports users in filtering results by their time and material constraints and then displays only those print resolution profiles and model scales that are within the time and material the user has available.

Finally, while SliceHub will contain increasing amounts of data over time, there will always be cases where either the desired print resolution profile and model scale have not yet been sliced or the 3D model is new to the shared repository and no data is available yet. For these cases, SliceHub provides infrastructure to either interpolate or slice the missing print resolution profiles and model scales in parallel. For parallel slicing, SliceHub interfaces with an existing slicing engine and instantiates multiple slicing processes with different print resolution profiles and model scales simultaneously. To make this scalable, SliceHub uses a cloud computing infrastructure that can return up to 1000~slicing results in parallel. SliceHub's parallel slicing infrastructure can thus return the slicing results for several hundred print resolution profiles and model scales combinations in the same time it would take to return results for only one. 
\vspace{12pt}

In summary, we contribute:
\begin{itemize}
    \item an augmented repository that stores for each 3D model the required print time and material consumption for different print resolution profiles and model scales, which allows future users to re-use the results in real-time;
    \item a user interface integrated into an existing slicer that supports exploration of multiple print resolution profiles and model scales at once, including functionality to narrow down available options by filtering based on time and material constraints;
    \item infrastructure to efficiently generate data for missing print resolution profiles and model scales  on the shared repository either through interpolation or by slicing multiple print resolution profiles and model scales in parallel.
\end{itemize}

\section{Related Work}
Our work is related to support tools for 3D printing, algorithms that improve the slicing process, research on finding speed-fidelity trade-offs, and data augmentations for 3D model repositories.

\subsection{Supporting Users with 3D Printing}
\cite{hudson2016understanding} were among the first to provide an in-depth analysis of the issues novice users (‘casual makers’) encounter when using 3D printers including the creation of the 3D~model itself and the subsequent slicing and fabrication process. Recent studies, including~\cite{dew2019producing, annett2019exploring, norouzi2021making} analyze fabrication workflows for users in various settings, ranging from makerspaces and fabrication studios \cite{annett2019exploring} to summer programs for young people \cite{norouzi2021making}.
In recent years, much of HCI research has focused on supporting users in the modeling process: Measurement Uncertainty~\cite{kim2017understanding}, for instance, compensates for users’ measurement errors in the initial modeling phase; RetroFab~\cite{ramakers2016retrofab} supports users in creating 3D models that add functionality to existing devices; and Lamello~\cite{savage2015lamello} enables users to create 3D designs with interactive input components. While there is a large body of work in HCI on improving 3D modeling tools as part of maker software~\cite{schmidt2010meshmixer}, only few research projects have focused on the later steps when a design is prepared for fabrication.
Most existing work on helping users prepare an object for fabrication has focused on laser cutting (VisiCut~\cite{oster2011visicut}, PacCam~\cite{saakes2013paccam}, Fabricaide~\cite{fabricaide2021}). When explored in the context of 3D printing existing work created new workflows not based of conventional tools. For instance, Scrappy~\cite{wall2021scrappy} suggests objects to be inserted into the 3D model to replace infill material. Our work, in contrast, builds onto slicing tools already in use today and can thus be directly integrated into the existing fabrication pipeline.

\subsection{Improved Slicing Algorithms}
Over the last years, several research projects in computer graphics have investigated how to improve slicing algorithms to improve print quality (Connected Fermat Spirals~\cite{zhao2016connected}, CurviSlicer~\cite{etienne2019curvislicer}), generate faster printing supports~\cite{schmidt2014branching, vanek2014clever, dumas2014bridging}, create better packing layouts on the print platform (Chopper~\cite{lu2014build}, Dapper~\cite{chen2015dapper}),  improve visual quality of the print result (Perceptual Support~\cite{zhang2015perceptual}, Saliency Preserving Slicing~\cite{wang2015saliency}), and increase durability (Cross-Section Analysis~\cite{umetani2013cross}, Orthogonal Slicing~\cite{hildebrand2013orthogonal}, Build-to-last~\cite{lu2014build}). Finally, researchers have also investigated how to create slicing algorithms that create new object properties, such as infills distributed in a way that makes an otherwise unbalanced object stand~\cite{prevost2013make} or spin~\cite{bacher2014spin} after fabrication. Rather than inventing new algorithms for slicers, our work investigates how to facilitate the exploration of slicing results for a 3D model.

 \subsection{Exploring Fabrication Trade-Offs}
Supporting users in finding the best trade-off between speed and fidelity has a long history in HCI research and has recently also been explored in the context of 3D printing. Low-fidelity fabrication techniques~\cite{mueller2015low}, such as WirePrint~\cite{mueller2014wireprint} and Platener~\cite{beyer2015platener}, for instance, allow users to trade-off model fidelity with print speed during the fabrication stage. During the modeling stage, SPATA~\cite{weichel2015spata} visualizes where support material will be generated, which allows designers to come to an informed decision about either modifying the design to avoid supports or spending the extra time on printing them. Our work also supports users in finding trade-offs between different options but applies it to the context of finding the most suitable print profile resolutions and model scales for 3D printing.

\subsection{Augmentations to 3D Model Repositories}
Several researchers explored how to augment existing 3D model repositories with additional data. Grafter~\cite{roumen2018grafter}, for instance, annotates mechanisms embedded in shared 3D models resulting in a model-graph of mechanisms that can be used for remixing mechanical parts. ShapeNet~\cite{shapenet2015} annotates 3D models with a rich set of geometric and language annotations that enable researchers to filter models via attributes like part decompositions and word taxonomies. Thingi10K~\cite{zhou2016thingi10k}, a dataset of 10,000 3D models, augments 3D models with several mesh complexity and quality metrics, such as the number of vertices and if the mesh is closed or self-intersects, which provides researchers with data that reflects real-world imperfect meshes instead of the clean data often used by researchers. ThingiPano~\cite{berman2020thingipano} is a large dataset, which includes 3-axis panoramic depth-map projections of 3D models together with images and meta-data, which can be used for machine learning. Finally, HowDIY~\cite{berman2021howdiy} provides a dataset and platform that facilitates the exploration of various online resources for 3D printing. In contrast to these datasets and platforms, SliceHub augments 3D models with their slicing results (print time, material consumption) for a variety of print profile resolutions and model scales to facilitate the exploration of slicing settings.
\section{Formative Study}

To better understand the challenges regarding slicing and the strategies employed by users to achieve various slicing goals for 3D printing, we conducted interviews with 18 participants (6f/12m). Participants  had used 3D printers for a diverse range of applications: such as 3D printing models for personal use, printing prototypes for research purposes, and for commercial and architectural use cases. We focused on topics such as the issues faced by participants when using slicers, types of print constraints faced while printing, experiences when exploring models on online 3D model repositories, and ideas for improvements. We identified several recurring break-downs.\\ 

\noindent\textbf{Long Wait Times}: Participants expressed frustration with the slow feedback of the slicer. Since the slicer by default re-slices the model after every print resolution profile and model scale change, participants stated that they often have to wait for the slicer even after making only minor changes. Participant P6, for instance, complained about the \textit{``long calculation times"} and P8 stated \textit{``every time I choose a profile, it needs a lot of time to calculate the print time."} Participants wondered if the wait time could be shortened. For instance, P6 stated \textit{"It would be cool if instead of calculating the entire slice operation after every change, the program would only updated time/material use and then just do a full slice calculation once I choose to save."} P8 added that estimates for the print time and material consumption may be enough for initial exploration: \textit{``I actually don't need the very precise time, I just need to know [...] it will take at least 10 hours."}\\

\noindent\textbf{Lack of Information about how Print Resolution and Model Scale relate to Print Time and Material Consumption:} Participants expressed frustration that there was no information on how different print resolution profile and model scale changes would impact the print time and material consumption. P1 recounted that, \textit{"[there was] no intuition or indication of what would affect time and material amount and by how much. Had to do ridiculous binary searching to find optimal value."} Multiple participants (P2, P5, P8, P10, and P11) indicated a preference towards having a recommendation or optimization system that would "\textit{compute the settings based on some desired quality heuristics}" (P10). P11 suggested that it would be useful "\textit{if I can input a target print time, or filament usage, and it can recommend settings for me, then I have a baseline to start with.}"\\ 

\noindent\textbf{Difficulty in Satisfying Print Constraints:} Most 3D printers used by participants were shared, and when faced with a deadline or during rapid prototyping, participants (P8, P10, P12) worked under stressful time and material constraints. P8 recounted: \textit{"I really, really needed to get that model done within 20 hours or else I [wasn't] going to make it to the deadline; I had to to find another way. [...] I spent around 20 minutes finding the correct solution so that I can get the thing done."} P10 added: \textit{"[if] that means sacrificing a lot of resolution to get it, it's better than not having a product in the end."}  Participants (P7, P8, P10, P12) also expressed that exploring different print resolution profiles and model scales took them a large number of iterations before arriving at a suitable solution. They shared ideas for improving the process of finding suitable print settings under time or material constraints. P7, for instance, suggested that \textit{``If the slicing program is really smart, then if I can type my purpose [goal], it may suggest some things"} and P8: \textit{``[if] there's a bar, and it will tell me if you scale to this, how much material is used."}  \\ 

\noindent\textbf{Lack of Print Information on Online Repositories:} Almost all participants had used 3D model repositories, such as \textit{Thingiverse}, when looking for 3D models. But participants often had to go back-and-forth between the online repository and the slicer because the repository did not include any print information. P10 stated \textit{"the website doesn't give you a lot of the things that the slicer software does."} Participants (P4, P5, P7, P8, P10, P11) mentioned that including print-specific information for models on these shared repositories would aid in determining if the model is suitable. P4, for instance, said \textit{``having more information on the platform for a specific model before having to download it and testing it out would be great."} P5 added: \textit{"you could have different size[s], so you could say with this particular size, this is the printing time [..]"}\\


\noindent To address the problem of long wait times, SliceHub's integrated system makes slicing results, i.e. the print time and material consumption for different print resolution profiles and model scales, available in real-time by storing slicing results and sharing them among all users. By making the print time and material amount of multiple print profile resolutions and model scales simultaneously visible in a user interface, it supports users in gaining an understanding of how they relate to each other. To support users in finding print profile resolutions and model scales that match a user's time and material constraints, SliceHub provides filter functionality that narrows down the options to only those that match the time and material the user has available. Finally, SliceHub embeds basic print information, such as the average print time and material consumption for a 3D model, on the 3D model repository page, which allows users to take this information into account already during the search for a suitable 3D model.

\section{SliceHub}
SliceHub is an integrated system that allows users to efficiently explore slicing results for a 3D~model. It consists of three components: (1)~a repository  of 3D models that contains for each 3D model the slicing results, i.e. print time and material consumption, for various print resolution profiles and model scales; (2)~a user interface integrated into an existing slicer, Ultimaker Cura~\cite{curaUltimaker}, that supports simultaneous exploration of multiple print resolution profiles and model scales as well as filtering based on time- and material constraints; (3)~a computational infrastructure to efficiently generate data for missing print resolution profiles and model scales either by estimating slicing results through interpolation or by slicing multiple print resolution profiles and model scales in parallel on an external cloud-based system. Together, these components allow users to explore trade-offs between different print resolution profiles and model scales with respect to the print time and material consumption for 3D printing a model. 

Note that SliceHub's components work based on print resolution profiles, i.e. a set of slicer settings (layer resolution, print speed, infill density etc.) that taken together print the model at a certain resolution. This is similar to existing slicers which offer predefined print resolution profiles (e.g. 'fast(0.2mm)', 'fine(0.06mm)'). Varying individual slicer settings, when not done carefully by an expert user, can lead to invalid prints. We thus exclude this option from SliceHub and only use predefined print resolution profiles whose settings were calibrated by the manufacturer of the 3D printer. 

In the next sections, we explain each of the three main components of SliceHub in more detail.


\subsection{Repository of Slicing Data}

SliceHub’s shared repository stores for each 3D model the slicing results for different print resolution profiles and model scales when printed on different 3D printers and with different materials.\\

\noindent\textbf{Content Stored in the Repository:} For each 3D model, SliceHub stores the 3D model geometry (.stl) and a meta-data file (.json) that contains for each print resolution profile and model scale the print time and material consumption for fabricating the 3D model. This allows the user to explore various print settings in real-time to find the trade-off between print time, material consumption, print resolution and model scale that best fits their needs. Storing print time and material consumption only requires little storage space (12 KB in total for one model with 16 different print resolution profiles and 16 different model scales, i.e. 256 combinations), which is a requirement to make the repository scaleable. We achieve this small storage overhead by discarding the print path (.gcode) since it does not contain information required to make a decision on the trade-offs discussed above and, additionally, an average .gcode file has an approximate file size of 10 MB per print resolution profile and model scale. Storing 256~.gcode files, one for each combination of the 16~print profile resolutions and 16~model scales, would require 2.5 GB on average and, thus, does not scale for large repositories. Instead, we only store the meta-data in a .json file. At 12 KB per model, even large data bases with thousands of models could store and process this additional data.\\

\begin{figure}[h]
    \centering
    \includegraphics[width=0.4\textwidth]{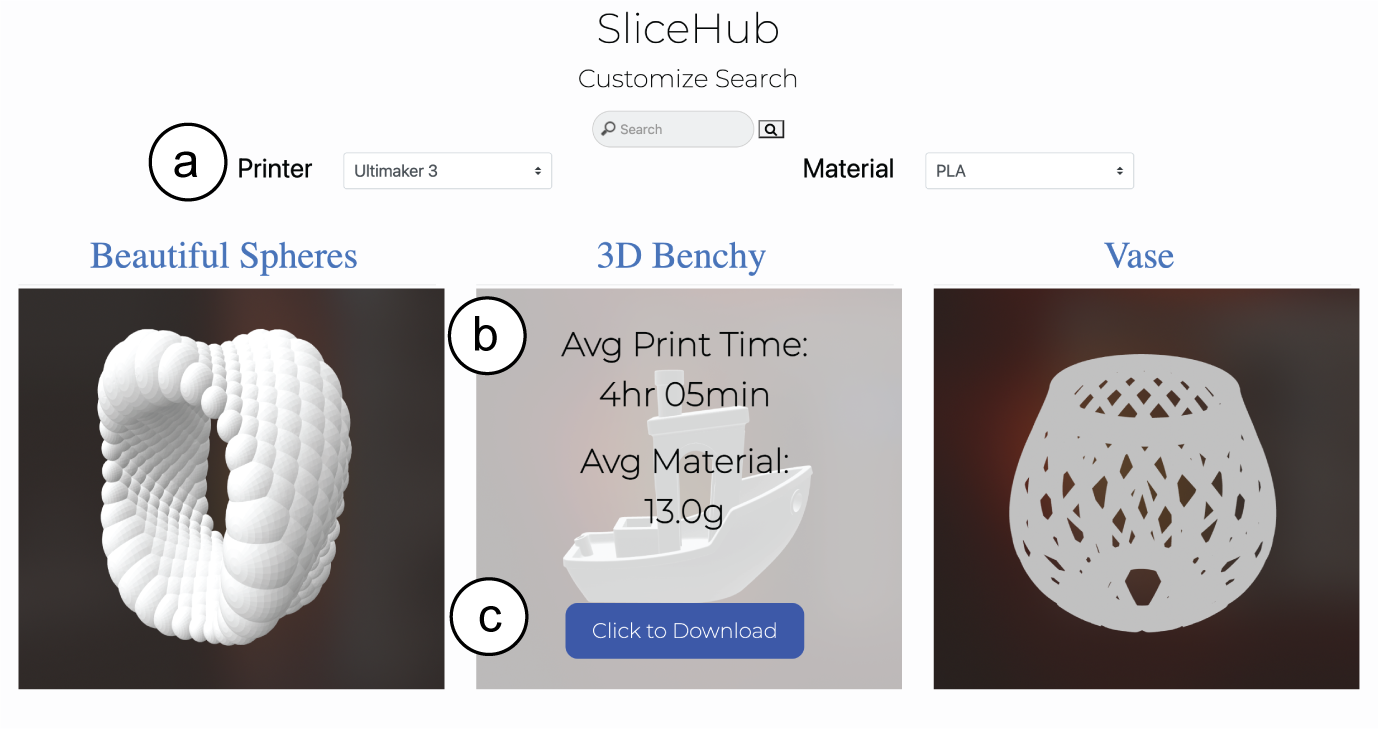}
    \caption{SliceHub repository: (a)~select 3D printer and material, (b)~for each model average print time and material consumption are shown, (c)~choose model and download .zip file (contains model geometry (.stl) and slicing results in meta-data file (.json)).} 
    \label{fig:2-slicehub-website}
\end{figure}

\noindent\textbf{Finding Content in the Repository:} The repository can be accessed through the SliceHub website and displays the 3D models on its main page (Figure~\ref{fig:2-slicehub-website}). The user starts by selecting the 3D~printer and material they want to use. Next, the user can start a query for specific models (e.g., "Mobius"). After clicking on the "Search" Button, SliceHub lists the most relevant models to the query in a result matrix. To provide the user with initial information on the print time and material consumption for a 3D~model, SliceHub shows this information at a medium print resolution profile (0.15mm) and at 100\% model scale when the user hovers over a 3D~model's thumbnail. In case no slicing results are available yet, the print time and material consumption are displayed as `not available'. Clicking on the 3D~model's thumbnail downloads a .zip archive containing the 3D model geometry (.stl) and a corresponding meta-data file (.json) that includes the slicing results for different print resolution profiles and model scales. As mentioned in the introduction, we envision that in the future the data from the SliceHub repository will be integrated with existing 3D~model platforms, such as \textit{Thingiverse}, with the print time and material consumption being made available during model search and on a 3D~model's subpage, and the meta-data file with slicing results added for download with the 3D model's .stl file.

\noindent\textbf{Adding Content to the Repository:} SliceHub’s repository can be populated with data in two ways. First, users have the option to `opt-in' for sharing their slicing results whenever they use SliceHub's parallel slicing infrastructure to generate new results (see Section \ref{sec:infrastructure-slicing-interpolation}). Second, SliceHub can generate slicing results itself using its cloud compute service and then subsequently add the results to the repository. Content can be added either to an existing model by providing slicing results for print resolution profiles and model scales that have no data yet or by contributing a new model that does not yet exist on the shared repository and adding an initial set of slicing results to it (see section `\ref{user-interface-tradeoffs} User Interface' for how users can add models). SliceHub identifies if a model already exists in the database or is new using the model's unique identifier (similar to \textit{Thingiverse}), which is also encoded in the meta-data file. Thus, once the slicing results are available in the repository, they can be reused by future users without the need to re-compute them.

\subsection{User Interface for Exploring Trade-Offs}
\label{user-interface-tradeoffs}
After downloading a 3D model and its meta-data file from the repository, users can load the content into the SliceHub user interface, which is integrated into an existing slicer (i.e., Ultimaker Cura). The user interface allows users to compare multiple print resolution profiles and model scales simultaneously, which facilitates the exploration of trade-offs. In addition, the user interface enables users to filter print resolution profiles and model scales according to material and time constraints.\\

\begin{figure}[h]
    \includegraphics[width=0.4\textwidth]{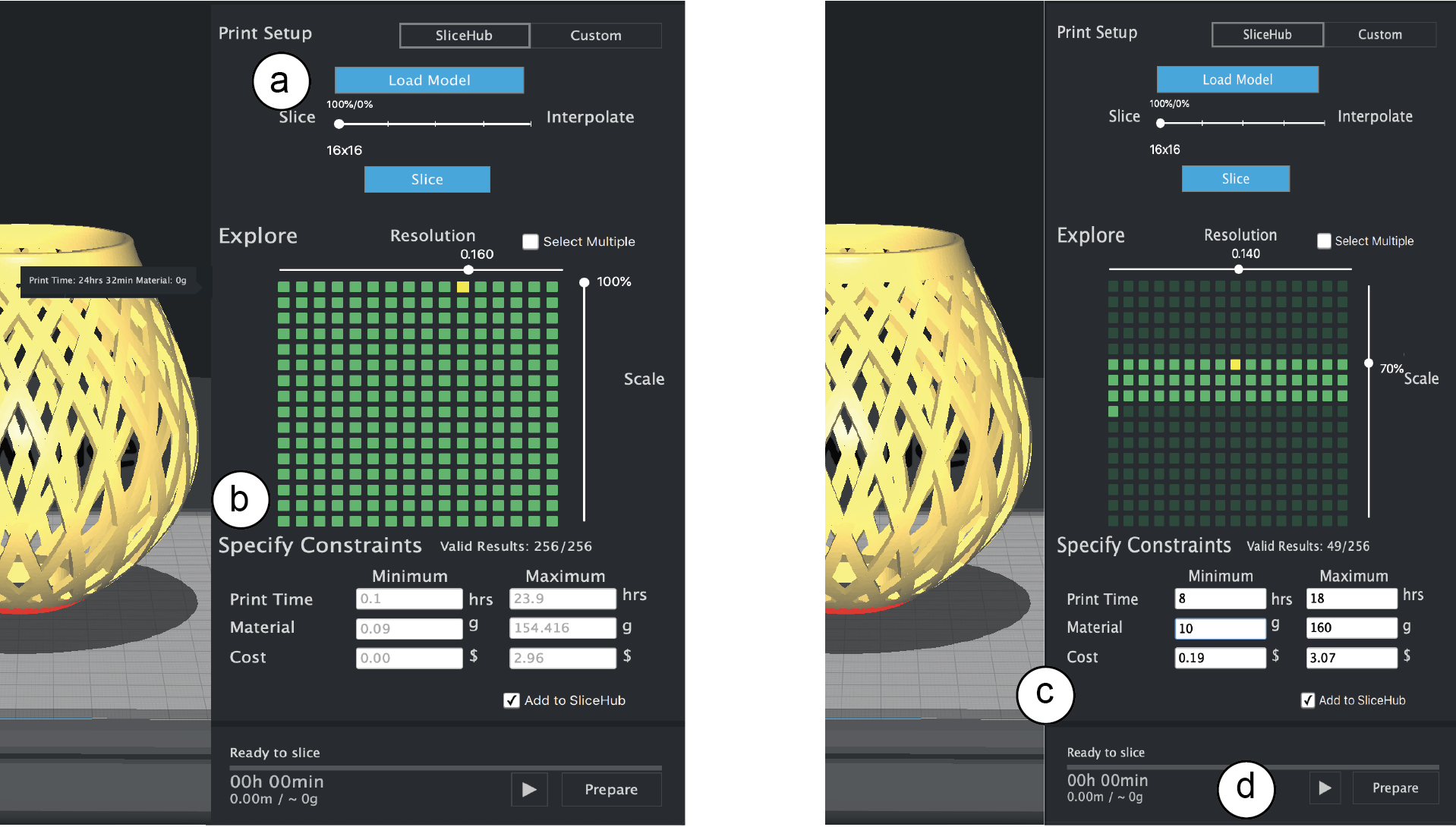}
    \caption{User Interface (Ultimaker Cura Plugin): (a)~load  .zip file (3D model (.stl) and meta-data (.json)), (b)~UI updates to display print time and material consumption for each print profile resolution and model scale. (c)~Adding lower and upper bounds for print time or material consumption shows only valid configurations. (d)~Slicing the preferred configuration to generate the .gcode file for 3D printing.}
    \label{fig:3-cura-user-interface}
\end{figure}

\noindent\textbf{Loading Data into the User Interface} The user interface takes as input the .zip file, which includes the 3D model geometry (.stl) and the meta-data file (.json) from the repository, which users can load using the `Load Model' button (Figure~\ref{fig:3-cura-user-interface}a). Next, SliceHub extracts the slicing results for different print resolution profiles and model scales from the meta-data file and populates the user interface with the corresponding print times and material amounts (Figure \ref{fig:3-cura-user-interface}b).\\

\noindent\textbf{Exploring Resolution and Scale Trade-Offs:} The visualization of slicing results (Figure \ref{fig:3-cura-user-interface}b) shows the print time and material amount for different print resolution profiles and model scales. Hovering over a slicing result displays the corresponding print time and material amount. The top left slicing result corresponds to the highest print resolution profile and largest model scale (0.06mm and 100\%) whereas the bottom right slicing result corresponds to the lowest print resolution profile and smallest model scale (0.2mm and 10\%). This allows users to compare different print profiles and model scales in real-time, which facilitates finding an appropriate trade-off. Since the focus of this paper is not on the particular visualization chosen to present the slicing results, we use a simple grid to represent the two-dimensional search space of print resolution profiles and model scales.\\

\noindent\textbf{Print Time/Material Constraints:} When 3D printing a model, users often work under time or material constraints as shown by our formative user study. For instance, users may not want to fabricate the model with a print resolution profile and model scale that requires printing overnight. SliceHub provides an option to narrow down the configurations to only those that print the model within the user’s available time and material.
Users can enter their constraints into the corresponding text boxes for lower and upper bounds (Figure~\ref{fig:3-cura-user-interface}c). To provide the user with information on the allowable range for these values, SliceHub initially populates the text boxes with the minimum and maximum values across all print resolution profiles and model scales for that particular 3D model (i.e., the print time and material consumption at the lowest resolution (0.2mm) and smallest scale (10\%), and highest resolution (0.06mm) and original model scale (100\%). As soon as the user enters the constraints, SliceHub updates the visualization, highlighting only those print resolution profiles and model scale combinations that are within the chosen bounds (Figure~\ref{fig:3-cura-user-interface}c). If users enter multiple constraints, such as both print time and material amount, both constraints are considered together.\\

\noindent\textbf{Generating GCode:} Once users settle on a print resolution profile and model scale combination by selecting it from the visualization, they can click the ‘Slice’ button (Figure~\ref{fig:3-cura-user-interface}d), which then slices the 3D model geometry (.stl) locally with the selected settings and afterwards loads the print path (.gcode) into the view. Users can then export the .gcode file using the existing slicer software functionality and then upload the .gcode file to their 3D printer to start the fabrication process. \\

\subsection{Infrastructure for Parallel Slicing and Interpolation}
\label{sec:infrastructure-slicing-interpolation}

SliceHub's infrastructure also addresses the need to generate new slicing results for print resolution profiles and model scales that have no data yet. To fill in the missing data, SliceHub can either interpolate the missing results, which can be done in real-time but is less accurate, or parallel slice the missing results, which is more accurate but requires additional time.\\ 

\begin{figure}[h]
    \includegraphics[scale=0.63]{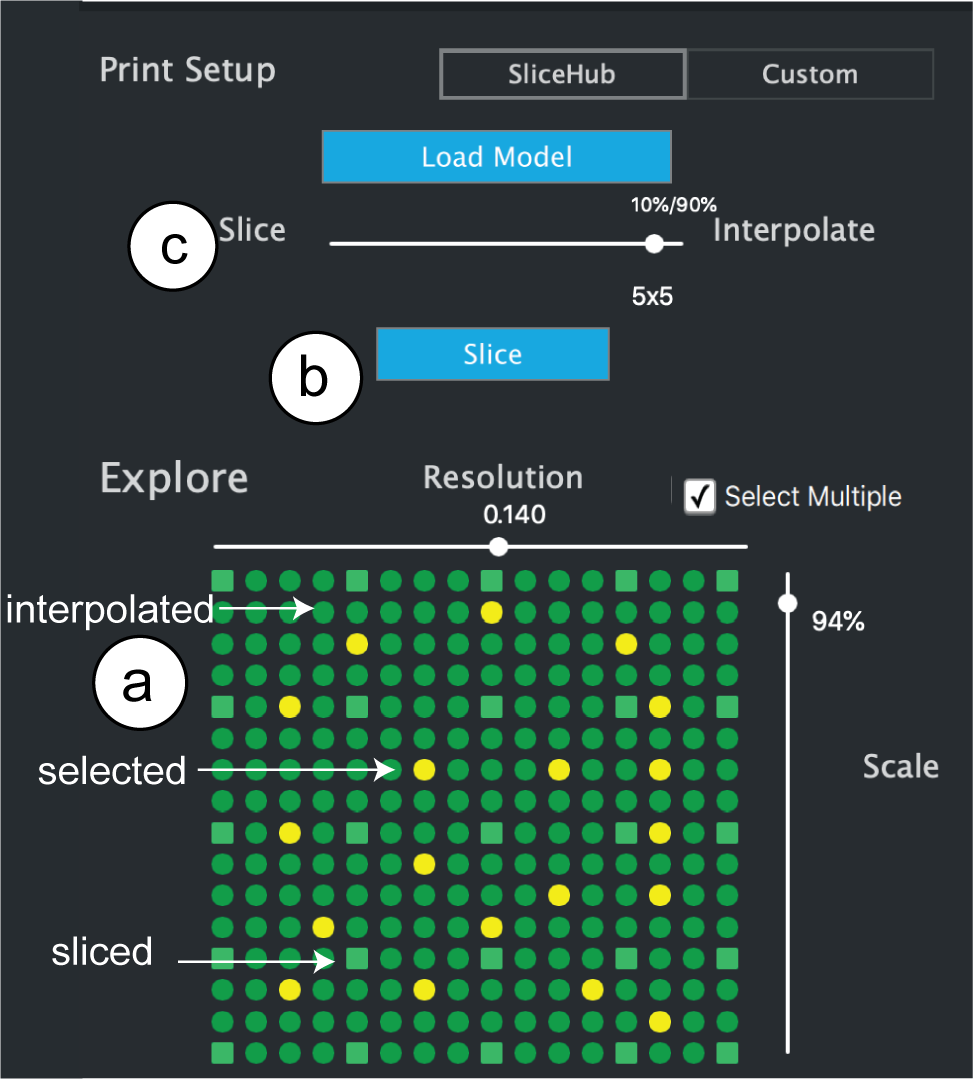}
    \caption{Adding slicing results: (a)~by default, all missing results are interpolated, (b)~selecting individual interpolated results and clicking `slice' turns them into sliced result, (c)~moving the `slice/interpolation' slider increases the percentage of sliced results via parallel slicing.}
    \label{fig:5-update-existing-results}
\end{figure}

\noindent\textbf{Interpolation of Missing Results:} When slicing results are not yet available, SliceHub interpolates the missing results from the already sliced results. To indicate which print resolution profile and model scale combinations are sliced and which are interpolated, SliceHub marks the sliced and interpolated results with different symbols in the visualization (Figure~\ref{fig:5-update-existing-results}). The interpolated values can be computed in real-time when the meta-data file (.json) is loaded. Thus, the interpolated results are immediately available for the user to explore. When the user hovers over one of these results, SliceHub shows the estimated print time and material consumption and also notes the prediction accuracy. \\

\begin{figure}[h]
    \includegraphics[width=0.3\textwidth]{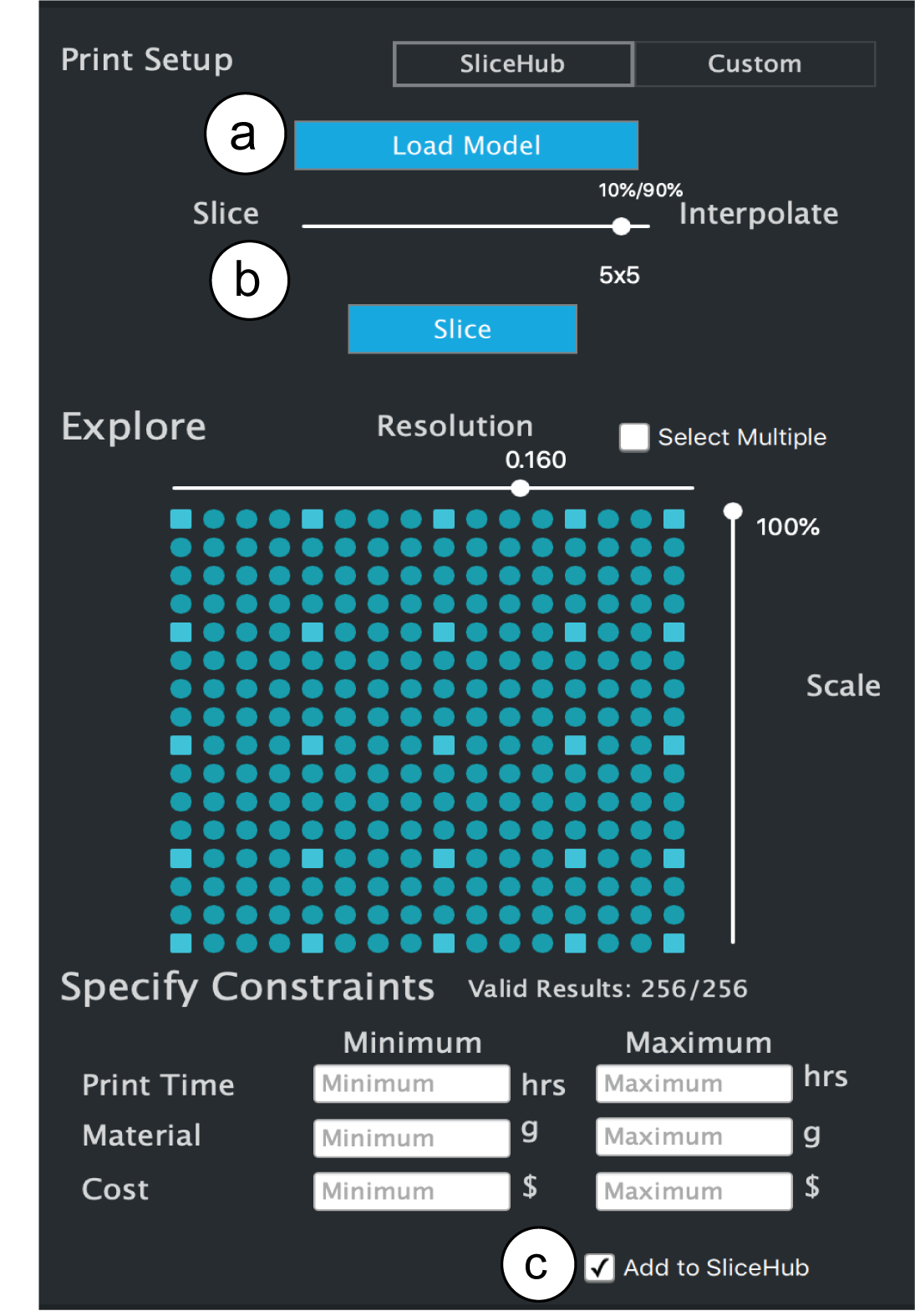}
    \caption{Adding a new model: (a)~load model geometry (.stl), (b)~SliceHub slices 10\% of the print profile and model scale combinations and interpolates the rest, (c)~SliceHub checks if users allowed sharing the model on the repository and then uploads the model and slicing results.}
    \label{fig:6-adding-new-model}
\end{figure}

\noindent\textbf{Parallel Slicing to Replace Interpolated Results:} To compute the slicing results for one or more interpolated values, users select the interpolated slicing results and then hit the ‘slice’ button (Figure~\ref{fig:5-update-existing-results}a). 
Alternatively, users can also use the ‘interpolation percentage’ slider (Figure~\ref{fig:5-update-existing-results}b), which allows to decrease the number of interpolated results and thus increase the accuracy of the prediction. To speed up the process of generating new slicing results, SliceHub contains infrastructure to compute slicing results for multiple print resolution profiles and model scales combinations in parallel. Note that when all slicing processes run in parallel, the total time equals the time for a single slicing process, i.e., the time required for the setting that needs the most slicing time, which is the model at 100\% scale printed with the highest print resolution profile. While the slicing results are being generated, the progress slider at the bottom of the interface updates to provide the user with an estimate of the remaining slicing time. Once the slicing results are available, the previously interpolated results update, i.e. convert their visual icon and on hover show the print time and material consumption for the sliced result rather than the interpolated estimate. When the `Add to SliceHub' option in the user interface is checked (Figure~\ref{fig:6-adding-new-model}c), all slicing results are added to the SliceHub repository for future re-use by other users (i.e., when SliceHub parallel slices in the cloud it stores the slicing results in the repository by updating the meta-data file of the model at the same time it returns the results to the user interface). However, users have the choice to override this setting by de-selecting the checkbox. The slicing results are then not stored on SliceHub and only saved as a new meta-data file on the user's local machine.\\

\noindent\textbf{Periodically Adding Results to Popular Models:} While we discussed in the previous sections how users can generate additional slicing results by selecting the desired print configurations in the user interface, SliceHub can also generate slicing results automatically through the repository. For instance, SliceHub has the capability to go over the models in its repository and identify those with missing slicing results. If SliceHub has processing capacity available, it generates the slicing results itself through its cloud compute service and adds them to the repository. If processing capacity is limited, SliceHub prioritizes popular models (by number of downloads) since the slicing results generated for those models will benefit the largest number of users on the shared repository. Once SliceHub added the missing results by slicing each print resolution profile and model scale combination once, they are available to all future users.\\

\noindent\textbf{Adding a New Model: Slicing + Interpolation:} To add a new 3D~model, SliceHub uses a combination of slicing and interpolation. Users start by loading the 3D model geometry (.stl) into the SliceHub user interface using the `Load Model' button (Figure~\ref{fig:6-adding-new-model}a), which instantiates an empty visualization with no available slicing results yet. If users have the `Add to SliceHub' option selected, SliceHub uploads the 3D model automatically to its cloud compute service and parallel slices it to generate 10\% of  the slicing results while interpolating the rest of the values (Figure~\ref{fig:6-adding-new-model}b). SliceHub uses 10\% since our technical evaluation shows that this fraction of sliced results leads to only a small interpolation error (Section~\ref{section:parallel-slicing}). SliceHub then saves the 3D model together with a newly generated meta-data file that contains the slicing results (.json) on the server before returning the results to the user interface. After this step, the 3D~model is available on the shared repository along with the generated slicing results. Users can decide to slice additional results, which are then added to the meta-data file on the server when SliceHub returns values from parallel slicing on its cloud compute service.

\section{Application Scenarios}
We next describe how SliceHub can be used for different applications.\\

\noindent\textbf{Exploring Model Scale vs. Print Resolution Trade-Offs:} In this scenario, an architect is short on time to print out a building model for an upcoming customer presentation. He loads the new 3D model into the SliceHub user interface and since the 3D model is not present in the repository, SliceHub proceeds to parallel slice 10\% of the data and interpolate the rest (see Technical Evaluation \ref{sec:interpolation-eval}). After looking at the SliceHub results for the highest print resolution profile and 100\% model scale, the architect concludes that printing for 26 hours is beyond the time he has available since the customer presentation is in 16 hours. The architect wants to preserve all the fine features of the building design and thus would like to print the model at the highest print resolution (0.06mm) if possible. He thus decides to explore if scaling down the model would print within his time constraint. However, upon closer inspection he realizes that if he kept the highest print resolution, he would have to scale the model by more than 50\%, which would make the model too small. He thus decides to slightly reduce print resolution to 0.1mm, which allows him to print the model at 80\% scale within 15 hours and 27 minutes. The architect slices the building 3D model locally with the selected model scale, downloads the gcode file, and starts printing.\\ 

\noindent\textbf{Exploring Models based on Print Times:} In this application scenario, a user wants to 3D print a flower pot for a plant they want to give to their friend for their birthday, which is the next day. They have limited access to a shared 3D~printer in the local library, and after looking at the open 3D printing times they see that for that particular day, there is only a 10~hour print window left. Since the user needs the flower pot within the day, she is searching for a model that prints within that time window. The user goes onto the SliceHub website and finds five different 3D models of flower pots, of which two have an average print time of under 10 hours. One of them looks more aesthetically pleasing, so the user decides to go with this model. She downloads the 3D model, opens it in the SliceHub user interface, generates the gcode for the default print settings that were shown on the repository, and sends the gcode file to the local library for printing. \\

\noindent\textbf{Exploring how to Reduce Material Usage:} In this application scenario, a maker is prototyping an interactive toy that has capacitive touch buttons across its surface. Since the conductive filament is expensive (Electrify Conductive Filament \cite{multi3d}, ca. \$200 per filament roll) , the maker wants to make sure that the prints during prototype iteration are not using too much material. The maker loads the 3D model into SliceHub, which causes SliceHub to parallel slice 10\% of the print profile resolutions and model scales. When exploring different model scales, the maker is surprised that scaling the model to 50\% saves more than 80\% of material since scaling down the model linearly leads to a cubic decrease in material volume. She thus decides that scaling the model to 75\% is sufficient for saving a lot of material. After printing the first version and verifying that the toy works as expected, the maker prints the toy at full scale. Since the maker is happy to share the 3D model and slicing results with others, she selects the 'Add to SliceHub' checkbox, which makes the model together with the slicing results available on the repository for other users to reuse. 


\section{Implementation}
SliceHub’s components, i.e. the repository, the user interface, and the infrastructure for parallel slicing and interpolation can be implemented in conjunction with a variety of existing cloud services and slicers. For the purposes of building a prototype system, we are using Amazon Web Services for the cloud storage and cloud slicing, and the slicer Ultimaker Cura \cite{curaUltimaker} for the user interface and slicer back-end. We provide details on the implementation of each of the components in the following sections.

\subsection{Repository Backend and Frontend}
There are three major components that make up the repository: (1)~a website built with ThreeJS \cite{danchilla2012three}, (2)~a cloud storage that contains all the data for the 3D models and their slicing results (AWS S3), and (3)~cloud processing infrastructure for parallel slicing consisting of a Flask \cite{grinberg2018flask} webserver connected to Amazon Web Services (AWS Lambda).
\vspace{10pt}

\noindent\textbf{Data Structure:} The cloud storage contains for every 3D model, the 3D model geometry (.stl file) and a meta-data file (.json) that contains for each print resolution profile and model scale the corresponding print time and material consumption. In addition, the cloud storage also saves a thumbnail for each model (generated from the .stl file using the numpy-stl library~\cite{oliphant2006guide}).
The web server also contains a global meta-data file (.json) that  contains links to each 3D model folder on the cloud storage.\\

\noindent\textbf{Data Exchange:} When the user hits the ‘download’ button, SliceHub packs both the 3D model geometry (.stl) and the meta-data file (.json) in a .zip container which can be opened in the SliceHub user interface embedded in the existing slicer Ultimaker Cura.\\

\noindent\textbf{Adding Data:} Every time SliceHub generates new slicing results, it updates the meta-data file of the 3D model on the cloud compute service. When a new 3D model is added, a new folder is created and the global meta-data file of the repository is updated with the new 3D model entry.

\subsection {User Interface Plugin}
The user interface is implemented as a plugin to Ultimaker Cura~3.6 using Cura’s plugin support. The backend is written in Python and the front-end is written in QML and Javascript. When users load the .zip file from the repository, our slicer plugin unpacks the file and after loading the 3D model (.stl) into the view, populates the user interface, i.e. adds to each print resolution profile and model scale the print time and material consumption from the meta-data file (.json). Next, it fills the constraint boxes for print time and material consumption with the respective minimum and maximum values from the print resolution profiles with the highest/lowest resolutions and largest/smallest model scales. All elements in the user interface are linked with each other, for instance, as users select a slicing result with a smaller model scale, the model is resized in the view.

\subsection{Parallel Slicing Infrastructure}
\label{section:parallel-slicing}

To be able to slice multiple print resolution profiles and model scales at the same time, we automated the slicing process and then deployed the automated slicer on the SliceHub infrastructure that enables us to run multiple slicing instances in parallel.\\ 

\noindent\textbf{Automated Slicing (CuraEngine):} For automated slicing, we use CuraEngine, an open source C++ library that can be operated as a command line tool. CuraEngine’s slice() function takes as input: (1) an .stl file for the 3D model, (2) a .json file containing the default print settings, and (3) a string containing the modified print parameters for each configuration (when the user selects a different print resolution profile the print settings, such as layer-height and print speed, change). After slicing, CuraEngine returns the print time and material amount to the terminal. SliceHub then reads the print time and material amount from the terminal and adds them to the model-specific meta-data file. Since SliceHub does not use the .gcode file that is also generated as a part of the slicing process, it discards it. Once the meta-data is updated, it can either be read back to the user interface embedded into the slicer or updated on the model repository website.\\

\noindent\textbf{Parallel Slicing in the Cloud:} To make slicing run in parallel, SliceHub uses the cloud-compute service Amazon Web Services and instantiates as many cloud computers in parallel as needed. Each cloud computer is instantiated with a .zip file that contains CuraEngine and a python script, which starts the slicing process as soon as the cloud computer is called. After slicing has finished, the print time and material are returned to the terminal by the cloud computer. The cloud computers then save the model-specific meta-data file on the cloud storage system. Depending on the print resolution profile and model scale, some slicing processes may finish faster than others. The user interface waits for all processes to return and then reads the model-specific meta-data file from the cloud storage to update the slicing results in the user interface. 

\vspace{10pt}

\subsection{Interpolation Algorithm}

To be able to estimate slicing results for missing print resolution profiles and model scales, we fit a function over the existing slicing results of a 3D model. When users increase the percentage of sliced results using the `interpolated/sliced' slider, we distribute sliced results uniformly across the space of print resolution profiles and model scales.\\ 

\noindent\textbf{Fitting a Function over Slicing Results:} To compute the print time and material consumption estimates for the interpolated slicing results, SliceHub fits a function to the existing slicing results of the currently loaded 3D model using the \textit{scikit-learn} library. SliceHub fits a second degree polynomial to all slicing results that are available for that particular model. We use a polynomial function since changes in print time and material consumption do not change linearly with changes in scale. The reason for this is that the overall volume of an object shows an approximately quadratic increase with a linear increase in scale. \\

\noindent\textbf{Choosing Which Print Profile Resolutions and Model Scales to Slice}: When users increase the number of sliced results using the `interpolated/sliced' slider, SliceHub computes which print resolution profiles and model scales should be sliced and which should remain interpolated. Given the selected percentage, SliceHub distributes the sliced results across the different print resolution profiles and model scales uniformly per dimension and uses the closest approximation when results cannot be uniformly distributed. 


\section{Technical Evaluation}
\label{sec:technical-evaluation}

We based our design decisions for the individual SliceHub components, such as what information to store in the repository and how many print profiles and model scales to provide in the user interface, on a set of technical evaluations, which we describe below.

\begin{figure*}[h]
    \includegraphics[width=\textwidth]{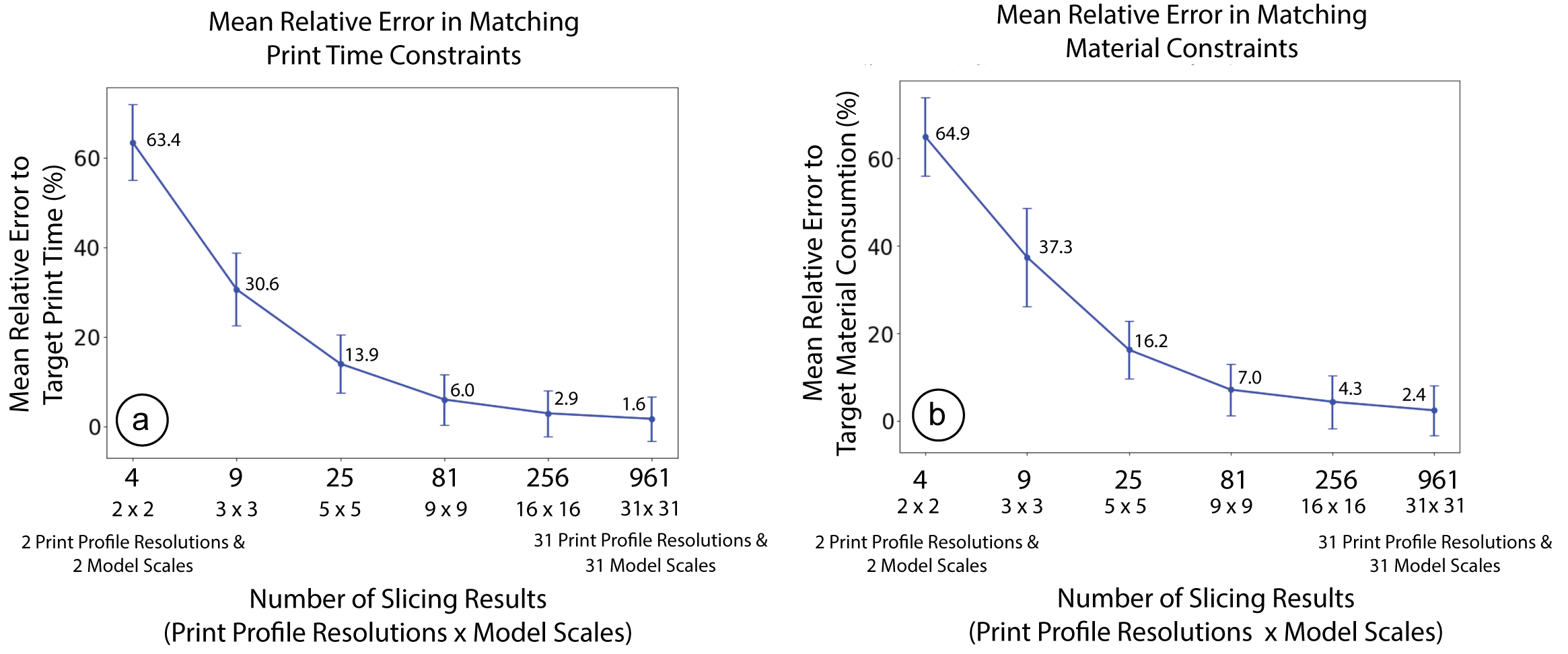}
    \caption{Mean relative error for different numbers of print resolution profiles and model scales offered in the user interface when either (a)~a print time constraint or (b)~a material amount constraint need to be matched.}
    \label{fig:9-fig-constraint}
\end{figure*}

\subsection{Repository Scalability: Amount of Meta-Data per Model}
\label{sec:repository-scalability}
We investigated how much storage space a model with all print resolution profiles and model scale combinations requires when all results are sliced and when a large fraction of results is interpolated.\\

\noindent\textbf{Average File Size All Results Sliced:} For the top 1200 Thingiverse models, we found that the meta-data file (.json) for one 3D~printer and material combination with 16x16 print resolution profiles and model scales (256 combinations) and with all results sliced is approximately 12 KB. When the top 7 materials are supported, which cover 90\% of the 3D printing demand in plastic~\cite{3DhubsTrendsReport}, the file size increases to approximately 84 KB. Further supporting the top ten 3D printers, which cover 28.4\% of all 3D printing users~\cite{3DhubsTrendsReport}, increases the file size to approximately 840 KB.\\

\noindent\textbf{Average File Size 90\% Results Interpolated:} When 90\% of results are interpolated as recommended by our technical evaluation 7.3, the average file size is reduced by 90\%, i.e. 1.2 KB rather than 12KB for one material/printer combination, 8.4 KB rather than 84KB for the top 7 materials, and 84 KB rather than 840KB for the top 10 printers.\\

\noindent\textbf{Average File Size of Geometry (.stl) and .gcode:} We found that for the top 1200 Thingiverse models, the average 3D model geometry file (.stl) was 6MB and the average  size for the .gcode file was 10MB.\\

\noindent Based on this data, we conclude that adding the SliceHub meta-data to an existing repository, such as \textit{Thingiverse,} for a range of popular print materials and 3D printers would only require an increase in the storage by 13.6\% if all results are sliced (840 KB on top of the 6MB average model geometry file size), or 1.36\% if 90\% of values are interpolated (84 KB on top of the 6MB average model geometry file size). Storing the gcode, however, would add significant storage overhead to the repository. Since we generate 1750 slicing results per model (10\% of 256 slicing results for 7 materials and 10 printers) and the size of an average gcode file is 10Mb, the overall storage increase would be approximately 17Gb per model (10Mb gcode * 1750 slicing results) and is thus not scale-able.

\subsection{User Interface: Effect of Number of Print Profiles on Constraint Accuracy}

When designing the user interface, we also had to decide on the number of print resolution profiles and model scales that the user can choose from. If there are only few options available, the user may not be able to find a good match for their time and material constraints. For instance, if only two print resolution profiles (e.g., 0.06mm and 0.2mm) are available, the high resolution profile may take much more time than the user has available while the fast low resolution may result in a worse print quality than the user would have had time for. Offering more options, however, increases computational cost to generate the slicing results. To find a good trade-off, we ran the following experiment.\\

\noindent\textbf{Conditions:} We created six conditions, i.e. six user interfaces with different numbers of print resolution profiles and model scales (ranging from 2 print resolution profiles and 2 model scales (4 slicing results) to 31~print profiles and 31~model scales (961 slicing results)). The user interface with the fewest options used the print resolution profiles with the highest resolution (0.06mm) and the lowest resolution (0.2mm), as well as the largest (100\%) and smallest (10\%) model scales. All other user interfaces iteratively added print resolution profiles and model scales at each midpoint of the existing values.\\ 

\begin{figure*}[h]
    \includegraphics[width=0.94\textwidth]{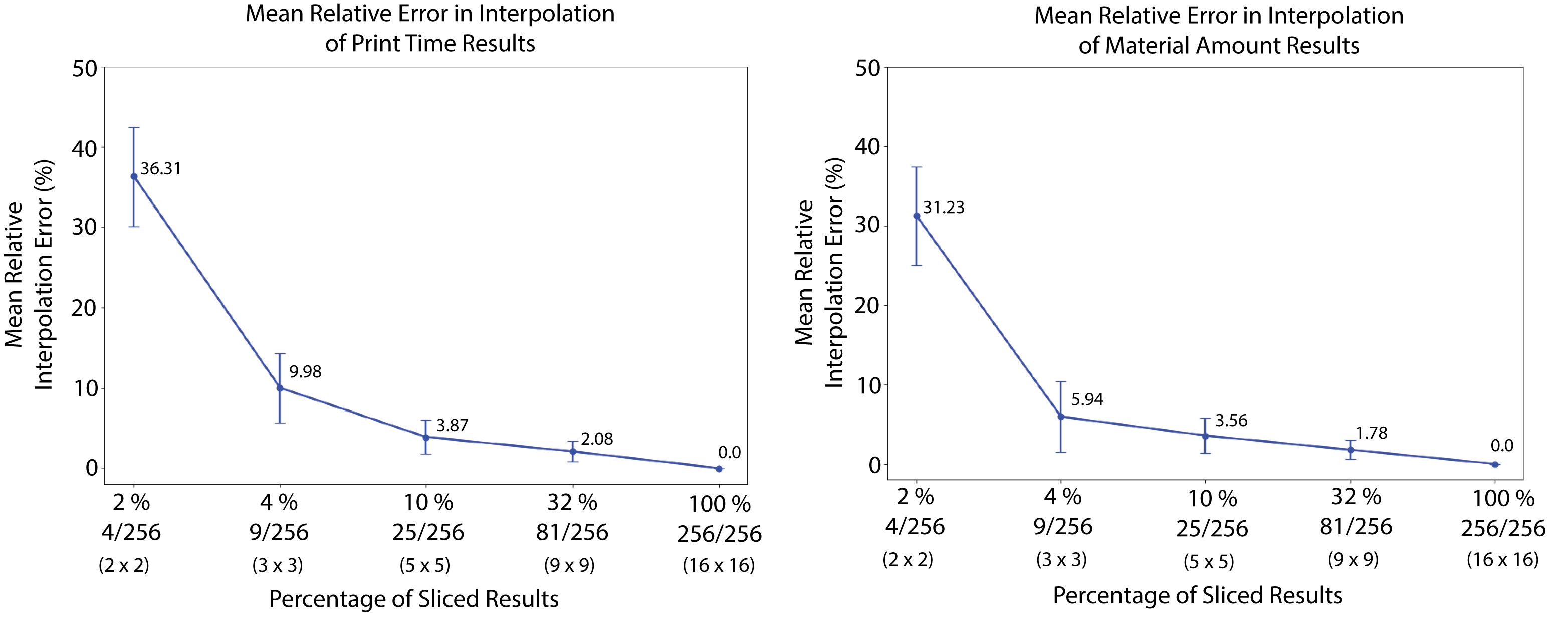}
    \caption{Higher percentages of interpolated results, lead to larger errors. The error stabilizes when 90\% of results are interpolated (i.e., 10
    \% of results are sliced). We thus use this as the default setting when a new model is added to SliceHub. }
    \label{fig:10-fig-prediction}
\end{figure*}

\noindent\textbf{Procedure:} We used our data-set of the top 1200 Thingiverse models and generated slicing results for each condition. We then randomly picked 20 print time and 20 material constraints from the available min/max bounds of each model, and for each user interface condition computed the mean relative error to the closest matching slicing result.\\

\noindent\textbf{Experiment Results:} As shown in Figure~\ref{fig:9-fig-constraint}, the more print resolution profiles and model scales the user interface offers, the smaller the mean relative error is. For the smallest user interface (2x2) the mean relative error for the print time is 63.4\% and material amount 64.9\%, whereas for the largest user interface (31x31) the mean relative error for print time is only 1.6\% and material amount 2.4\%. However, while there is a strong reduction in error when a small user interface is expanded (e.g., for print time: 2x2: 63.4\% vs 3x3: 30.6\%), the effect is less emphasized for larger user interfaces (e.g., for print time there is only a 1.3\% improvement between 16x16: 2.9\% vs. 31x31: 1.6\%). Since a user interface with 16 print resolution profiles and model scales requires only 26.6\% of the data of a user interface with 31~print resolution profiles and model scales, and thus significantly lowers computational cost, we decided to use the user interface with 16 print resolution profiles and model scales as the default for SliceHub.

\subsection{Effect of Number of Interpolated Slicing Results on Average Prediction Error}
\label{sec:interpolation-eval}

To decide how many slicing results SliceHub should by default slice and how many interpolate, we ran an experiment to evaluate the effect of an increasing number of interpolated results on the mean relative error for predicting print time and material consumption.\\

\noindent\textbf{Experiment Procedure:} We created five conditions, i.e. five user interfaces of 16x16 print resolution profiles and model scales that had different percentages of interpolated results. The user interfaces ranged from containing 0\% interpolated results (all 16 print resolution profiles x 16 model scales combinations sliced) to 98\% interpolated cells (only 2 print resolution profiles x 2 model scales sliced). We then calculated the relative mean error for the interpolated results from the computed slicing results (e.g., if the computed print time of an object is 2h but the interpolated print time is 1h, there is a relative error of 50\%). \\

\noindent\textbf{Experiment Results:} As can be seen in Figure~\ref{fig:10-fig-prediction}, the error in predicted print time and material amount decreases with number of sliced results since there are more data points available for accurately fitting the polynomial function. For instance, the interpolation error for print time when only 4 (2x2) out of 256 print resolution profiles and model scales are sliced is 36.31\%, whereas the interpolation error is only 2.08\% when 81 (9x9) out of 256 print resolution profiles and model scales are sliced. Comparing the interpolation error for different percentages of interpolated results shows a strong decrease in error when 98\% vs 96\% of results are interpolated (26.33\% less interpolation error) and when 96\% vs 90\% are interpolated (6.31\% less interpolation error). The reduction in interpolation error when 90\% vs. 68\% of results are interpolated is only marginal (1.79\% less error). Since interpolation accuracy no longer strongly decreases after about 10\% of sliced results (90\% interpolation), SliceHub by default slices 10\% of results when a new model is added.

\vspace{10pt}

\section{Discussion and Future Work}

We next discuss how SliceHub can be extended in the future.\\

\noindent\textit{Indicating Failed Print Settings:} While in our work, we have focused on making slicing results, such as print time and material consumption, readily available for exploration and comparison, we currently do not provide users with information if the chosen print profile resolutions and model scales lead to a successful print. For future work, we plan to allow the model designer or a user who printed the model to indicate if specific print profile resolutions and model scales worked or if the print did not complete. By integrating this information in the user interface, SliceHub can provide the information in a structured manner rather than through the informal comments section as is commonly done in 3D model repositories, such as Thingiverse.\\

\noindent\textit{Comparing Printers / Materials:} While we focused on comparing different print resolution profiles and model scales of a 3D model, the data in SliceHub's repository can also be used for other use cases. For instance, SliceHub's data allows users to compare the print time of different 3D printers with each other enabling users to draw conclusions which 3D printer on the market is currently the fastest. Similarly, users can compare the print times of different filament types with each other, which may influence their choice of print material.\\

\noindent\textit{Extending the Slicing Infrastructure:} Finally, we plan to extend SliceHub's slicing infrastructure. Our current implementation is based on the slicer Ultimaker Cura and thus only works for FDM 3D printers. By default, it is setup to generate slicing results for the Ultimaker 3D printer series but print resolution profiles for other FDM 3D printers (e.g., MakerBot, Prusa) can be imported and subsequently used for slicing. To support other 3D printing technologies, we are currently in the process of extending our slicing infrastructure with the slicer Slic3r \cite{slic3r}, which supports a wider variety of other 3D printers, such as those based on DLP printing (e.g., Prusa SL1 and Formlabs Form 3). 
\section{Conclusion}

We presented SliceHub, an integrated system that enables users to explore different print settings in real-time to find the trade-off between print resolution profile, model scale, print time, and material consumption that best fits their needs. We discussed the design and technical implementation of the SliceHub repository for storing slicing results, the user interface for exploring trade-offs between multiple print resolution profiles and model scales simultaneously, and SliceHub's infrastructure for slicing and interpolation that further adds to the data available in the repository. We provided an evaluation of the data storage requirements for the repository and reported evaluation results to support design decisions made in the user interface, such as the number of print resolution profiles and model scales the user can choose from. For future work, we plan to deploy the repository as a website and provide the plugin for the slicer through the Ultimaker Cura marketplace. In addition, we plan to track contributions to the repository and investigate users’ activities on SliceHub as well as collect feedback from the maker community.

\bibliographystyle{ACM-Reference-Format}
\bibliography{references}


\begin{thebibliography}{43}


\ifx \showCODEN    \undefined \def \showCODEN     #1{\unskip}     \fi
\ifx \showDOI      \undefined \def \showDOI       #1{#1}\fi
\ifx \showISBNx    \undefined \def \showISBNx     #1{\unskip}     \fi
\ifx \showISBNxiii \undefined \def \showISBNxiii  #1{\unskip}     \fi
\ifx \showISSN     \undefined \def \showISSN      #1{\unskip}     \fi
\ifx \showLCCN     \undefined \def \showLCCN      #1{\unskip}     \fi
\ifx \shownote     \undefined \def \shownote      #1{#1}          \fi
\ifx \showarticletitle \undefined \def \showarticletitle #1{#1}   \fi
\ifx \showURL      \undefined \def \showURL       {\relax}        \fi
\providecommand\bibfield[2]{#2}
\providecommand\bibinfo[2]{#2}
\providecommand\natexlab[1]{#1}
\providecommand\showeprint[2][]{arXiv:#2}

\bibitem[\protect\citeauthoryear{Annett, Grossman, Wigdor, and
  Fitzmaurice}{Annett et~al\mbox{.}}{2019}]%
        {annett2019exploring}
\bibfield{author}{\bibinfo{person}{Michelle Annett}, \bibinfo{person}{Tovi
  Grossman}, \bibinfo{person}{Daniel Wigdor}, {and} \bibinfo{person}{George
  Fitzmaurice}.} \bibinfo{year}{2019}\natexlab{}.
\newblock \showarticletitle{Exploring and understanding the role of workshop
  environments in personal fabrication processes}.
\newblock \bibinfo{journal}{\emph{ACM Transactions on Computer-Human
  Interaction (TOCHI)}} \bibinfo{volume}{26}, \bibinfo{number}{2}
  (\bibinfo{year}{2019}), \bibinfo{pages}{1--43}.
\newblock


\bibitem[\protect\citeauthoryear{B{\"a}cher, Whiting, Bickel, and
  Sorkine-Hornung}{B{\"a}cher et~al\mbox{.}}{2014}]%
        {bacher2014spin}
\bibfield{author}{\bibinfo{person}{Moritz B{\"a}cher}, \bibinfo{person}{Emily
  Whiting}, \bibinfo{person}{Bernd Bickel}, {and} \bibinfo{person}{Olga
  Sorkine-Hornung}.} \bibinfo{year}{2014}\natexlab{}.
\newblock \showarticletitle{Spin-it: optimizing moment of inertia for spinnable
  objects}.
\newblock \bibinfo{journal}{\emph{ACM Transactions on Graphics (TOG)}}
  \bibinfo{volume}{33}, \bibinfo{number}{4} (\bibinfo{year}{2014}),
  \bibinfo{pages}{1--10}.
\newblock


\bibitem[\protect\citeauthoryear{Berman and Quek}{Berman and Quek}{2020}]%
        {berman2020thingipano}
\bibfield{author}{\bibinfo{person}{Alexander Berman} {and}
  \bibinfo{person}{Francis Quek}.} \bibinfo{year}{2020}\natexlab{}.
\newblock \showarticletitle{ThingiPano: A Large-Scale Dataset of 3D Printing
  Metadata, Images, and Panoramic Renderings for Exploring Design Reuse}. In
  \bibinfo{booktitle}{\emph{2020 IEEE Sixth International Conference on
  Multimedia Big Data (BigMM)}}. IEEE, \bibinfo{pages}{18--27}.
\newblock


\bibitem[\protect\citeauthoryear{Berman, Thakare, Howell, Quek, and Kim}{Berman
  et~al\mbox{.}}{2021}]%
        {berman2021howdiy}
\bibfield{author}{\bibinfo{person}{Alexander Berman}, \bibinfo{person}{Ketan
  Thakare}, \bibinfo{person}{Joshua Howell}, \bibinfo{person}{Francis Quek},
  {and} \bibinfo{person}{Jeeeun Kim}.} \bibinfo{year}{2021}\natexlab{}.
\newblock \showarticletitle{HowDIY: Towards Meta-Design Tools to Support Anyone
  to 3D Print Anywhere}. In \bibinfo{booktitle}{\emph{26th International
  Conference on Intelligent User Interfaces}}. \bibinfo{pages}{491--503}.
\newblock


\bibitem[\protect\citeauthoryear{Beyer, Gurevich, Mueller, Chen, and
  Baudisch}{Beyer et~al\mbox{.}}{2015}]%
        {beyer2015platener}
\bibfield{author}{\bibinfo{person}{Dustin Beyer}, \bibinfo{person}{Serafima
  Gurevich}, \bibinfo{person}{Stefanie Mueller}, \bibinfo{person}{Hsiang-Ting
  Chen}, {and} \bibinfo{person}{Patrick Baudisch}.}
  \bibinfo{year}{2015}\natexlab{}.
\newblock \showarticletitle{Platener: Low-fidelity fabrication of 3D objects by
  substituting 3D print with laser-cut plates}. In
  \bibinfo{booktitle}{\emph{Proceedings of the 33rd Annual ACM Conference on
  Human Factors in Computing Systems}}. \bibinfo{pages}{1799--1806}.
\newblock


\bibitem[\protect\citeauthoryear{Chang, Funkhouser, Guibas, Hanrahan, Huang,
  Li, Savarese, Savva, Song, Su, Xiao, Yi, and Yu}{Chang et~al\mbox{.}}{2015}]%
        {shapenet2015}
\bibfield{author}{\bibinfo{person}{Angel~X. Chang}, \bibinfo{person}{Thomas
  Funkhouser}, \bibinfo{person}{Leonidas Guibas}, \bibinfo{person}{Pat
  Hanrahan}, \bibinfo{person}{Qixing Huang}, \bibinfo{person}{Zimo Li},
  \bibinfo{person}{Silvio Savarese}, \bibinfo{person}{Manolis Savva},
  \bibinfo{person}{Shuran Song}, \bibinfo{person}{Hao Su},
  \bibinfo{person}{Jianxiong Xiao}, \bibinfo{person}{Li Yi}, {and}
  \bibinfo{person}{Fisher Yu}.} \bibinfo{year}{2015}\natexlab{}.
\newblock \bibinfo{booktitle}{\emph{{ShapeNet: An Information-Rich 3D Model
  Repository}}}.
\newblock \bibinfo{type}{{T}echnical {R}eport} arXiv:1512.03012 [cs.GR].
  \bibinfo{institution}{Stanford University --- Princeton University --- Toyota
  Technological Institute at Chicago}.
\newblock


\bibitem[\protect\citeauthoryear{Chen, Zhang, Lin, Hu, Lu, Huang, Benes,
  Cohen-Or, and Chen}{Chen et~al\mbox{.}}{2015}]%
        {chen2015dapper}
\bibfield{author}{\bibinfo{person}{Xuelin Chen}, \bibinfo{person}{Hao Zhang},
  \bibinfo{person}{Jinjie Lin}, \bibinfo{person}{Ruizhen Hu},
  \bibinfo{person}{Lin Lu}, \bibinfo{person}{Qi-Xing Huang},
  \bibinfo{person}{Bedrich Benes}, \bibinfo{person}{Daniel Cohen-Or}, {and}
  \bibinfo{person}{Baoquan Chen}.} \bibinfo{year}{2015}\natexlab{}.
\newblock \showarticletitle{Dapper: decompose-and-pack for 3D printing.}
\newblock \bibinfo{journal}{\emph{ACM Trans. Graph.}} \bibinfo{volume}{34},
  \bibinfo{number}{6} (\bibinfo{year}{2015}), \bibinfo{pages}{213--1}.
\newblock


\bibitem[\protect\citeauthoryear{Danchilla}{Danchilla}{2012}]%
        {danchilla2012three}
\bibfield{author}{\bibinfo{person}{Brian Danchilla}.}
  \bibinfo{year}{2012}\natexlab{}.
\newblock \showarticletitle{Three. js framework}.
\newblock In \bibinfo{booktitle}{\emph{Beginning WebGL for HTML5}}.
  \bibinfo{publisher}{Springer}, \bibinfo{pages}{173--203}.
\newblock


\bibitem[\protect\citeauthoryear{Dew, Landwehr-Sydow, Rosner, Thayer, and
  Jonsson}{Dew et~al\mbox{.}}{2019}]%
        {dew2019producing}
\bibfield{author}{\bibinfo{person}{Kristin~N Dew}, \bibinfo{person}{Sophie
  Landwehr-Sydow}, \bibinfo{person}{Daniela~K Rosner}, \bibinfo{person}{Alex
  Thayer}, {and} \bibinfo{person}{Martin Jonsson}.}
  \bibinfo{year}{2019}\natexlab{}.
\newblock \showarticletitle{Producing printability: Articulation work and
  alignment in 3d printing}.
\newblock \bibinfo{journal}{\emph{Human--Computer Interaction}}
  \bibinfo{volume}{34}, \bibinfo{number}{5-6} (\bibinfo{year}{2019}),
  \bibinfo{pages}{433--469}.
\newblock


\bibitem[\protect\citeauthoryear{Dumas, Hergel, and Lefebvre}{Dumas
  et~al\mbox{.}}{2014}]%
        {dumas2014bridging}
\bibfield{author}{\bibinfo{person}{J{\'e}r{\'e}mie Dumas},
  \bibinfo{person}{Jean Hergel}, {and} \bibinfo{person}{Sylvain Lefebvre}.}
  \bibinfo{year}{2014}\natexlab{}.
\newblock \showarticletitle{Bridging the gap: automated steady scaffoldings for
  3D printing}.
\newblock \bibinfo{journal}{\emph{ACM Transactions on Graphics (TOG)}}
  \bibinfo{volume}{33}, \bibinfo{number}{4} (\bibinfo{year}{2014}),
  \bibinfo{pages}{1--10}.
\newblock


\bibitem[\protect\citeauthoryear{Etienne, Ray, Panozzo, Hornus, Wang,
  Mart{\'\i}nez, McMains, Alexa, Wyvill, and Lefebvre}{Etienne
  et~al\mbox{.}}{2019}]%
        {etienne2019curvislicer}
\bibfield{author}{\bibinfo{person}{Jimmy Etienne}, \bibinfo{person}{Nicolas
  Ray}, \bibinfo{person}{Daniele Panozzo}, \bibinfo{person}{Samuel Hornus},
  \bibinfo{person}{Charlie~CL Wang}, \bibinfo{person}{Jon{\`a}s Mart{\'\i}nez},
  \bibinfo{person}{Sara McMains}, \bibinfo{person}{Marc Alexa},
  \bibinfo{person}{Brian Wyvill}, {and} \bibinfo{person}{Sylvain Lefebvre}.}
  \bibinfo{year}{2019}\natexlab{}.
\newblock \showarticletitle{CurviSlicer: Slightly curved slicing for 3-axis
  printers}.
\newblock \bibinfo{journal}{\emph{ACM Transactions on Graphics (TOG)}}
  \bibinfo{volume}{38}, \bibinfo{number}{4} (\bibinfo{year}{2019}),
  \bibinfo{pages}{1--11}.
\newblock


\bibitem[\protect\citeauthoryear{Grinberg}{Grinberg}{2018}]%
        {grinberg2018flask}
\bibfield{author}{\bibinfo{person}{Miguel Grinberg}.}
  \bibinfo{year}{2018}\natexlab{}.
\newblock \bibinfo{booktitle}{\emph{Flask web development: developing web
  applications with python}}.
\newblock \bibinfo{publisher}{" O'Reilly Media, Inc."}.
\newblock


\bibitem[\protect\citeauthoryear{Hildebrand, Bickel, and Alexa}{Hildebrand
  et~al\mbox{.}}{2013}]%
        {hildebrand2013orthogonal}
\bibfield{author}{\bibinfo{person}{Kristian Hildebrand}, \bibinfo{person}{Bernd
  Bickel}, {and} \bibinfo{person}{Marc Alexa}.}
  \bibinfo{year}{2013}\natexlab{}.
\newblock \showarticletitle{Orthogonal slicing for additive manufacturing}.
\newblock \bibinfo{journal}{\emph{Computers \& Graphics}} \bibinfo{volume}{37},
  \bibinfo{number}{6} (\bibinfo{year}{2013}), \bibinfo{pages}{669--675}.
\newblock


\bibitem[\protect\citeauthoryear{Hubs}{Hubs}{2020}]%
        {3DhubsTrendsReport}
\bibfield{author}{\bibinfo{person}{3D Hubs}.} \bibinfo{year}{2020}\natexlab{}.
\newblock \bibinfo{title}{3D printing Trends 2020 Report}.
\newblock
\newblock
\urldef\tempurl%
\url{https://www.3dhubs.com/get/trends/}
\showURL{%
\tempurl}


\bibitem[\protect\citeauthoryear{Hudson, Alcock, and Chilana}{Hudson
  et~al\mbox{.}}{2016}]%
        {hudson2016understanding}
\bibfield{author}{\bibinfo{person}{Nathaniel Hudson}, \bibinfo{person}{Celena
  Alcock}, {and} \bibinfo{person}{Parmit~K Chilana}.}
  \bibinfo{year}{2016}\natexlab{}.
\newblock \showarticletitle{Understanding newcomers to 3D printing:
  Motivations, workflows, and barriers of casual makers}. In
  \bibinfo{booktitle}{\emph{Proceedings of the 2016 CHI Conference on Human
  Factors in Computing Systems}}. \bibinfo{pages}{384--396}.
\newblock


\bibitem[\protect\citeauthoryear{Kim, Guo, Yeh, Hudson, and Mankoff}{Kim
  et~al\mbox{.}}{2017}]%
        {kim2017understanding}
\bibfield{author}{\bibinfo{person}{Jeeeun Kim}, \bibinfo{person}{Anhong Guo},
  \bibinfo{person}{Tom Yeh}, \bibinfo{person}{Scott~E Hudson}, {and}
  \bibinfo{person}{Jennifer Mankoff}.} \bibinfo{year}{2017}\natexlab{}.
\newblock \showarticletitle{Understanding uncertainty in measurement and
  accommodating its impact in 3D modeling and printing}. In
  \bibinfo{booktitle}{\emph{Proceedings of the 2017 Conference on Designing
  Interactive Systems}}. \bibinfo{pages}{1067--1078}.
\newblock


\bibitem[\protect\citeauthoryear{Lu, Sharf, Zhao, Wei, Fan, Chen, Savoye, Tu,
  Cohen-Or, and Chen}{Lu et~al\mbox{.}}{2014}]%
        {lu2014build}
\bibfield{author}{\bibinfo{person}{Lin Lu}, \bibinfo{person}{Andrei Sharf},
  \bibinfo{person}{Haisen Zhao}, \bibinfo{person}{Yuan Wei},
  \bibinfo{person}{Qingnan Fan}, \bibinfo{person}{Xuelin Chen},
  \bibinfo{person}{Yann Savoye}, \bibinfo{person}{Changhe Tu},
  \bibinfo{person}{Daniel Cohen-Or}, {and} \bibinfo{person}{Baoquan Chen}.}
  \bibinfo{year}{2014}\natexlab{}.
\newblock \showarticletitle{Build-to-last: strength to weight 3D printed
  objects}.
\newblock \bibinfo{journal}{\emph{ACM Transactions on Graphics (TOG)}}
  \bibinfo{volume}{33}, \bibinfo{number}{4} (\bibinfo{year}{2014}),
  \bibinfo{pages}{1--10}.
\newblock


\bibitem[\protect\citeauthoryear{Ludwig, Stickel, Boden, and Pipek}{Ludwig
  et~al\mbox{.}}{2014}]%
        {ludwig2014towards}
\bibfield{author}{\bibinfo{person}{Thomas Ludwig}, \bibinfo{person}{Oliver
  Stickel}, \bibinfo{person}{Alexander Boden}, {and} \bibinfo{person}{Volkmar
  Pipek}.} \bibinfo{year}{2014}\natexlab{}.
\newblock \showarticletitle{Towards sociable technologies: an empirical study
  on designing appropriation infrastructures for 3D printing}. In
  \bibinfo{booktitle}{\emph{Proceedings of the 2014 conference on Designing
  interactive systems}}. \bibinfo{pages}{835--844}.
\newblock


\bibitem[\protect\citeauthoryear{Makerbot}{Makerbot}{2015}]%
        {makerbot}
\bibfield{author}{\bibinfo{person}{Makerbot}.} \bibinfo{year}{2015}\natexlab{}.
\newblock \bibinfo{title}{Celebrating a Maker Milestone: 1 Million Uploads on
  Makerbot's Thingiverse}.
\newblock
\newblock
\urldef\tempurl%
\url{https://www.makerbot.com/stories/2015/10/29/}
\showURL{%
\tempurl}


\bibitem[\protect\citeauthoryear{Mueller, Beyer, Mohr, Gurevich, Teibrich,
  Pfistere, Guenther, Frohnhofen, Chen, Baudisch, et~al\mbox{.}}{Mueller
  et~al\mbox{.}}{2015}]%
        {mueller2015low}
\bibfield{author}{\bibinfo{person}{Stefanie Mueller}, \bibinfo{person}{Dustin
  Beyer}, \bibinfo{person}{Tobias Mohr}, \bibinfo{person}{Serafima Gurevich},
  \bibinfo{person}{Alexander Teibrich}, \bibinfo{person}{Lisa Pfistere},
  \bibinfo{person}{Kerstin Guenther}, \bibinfo{person}{Johannes Frohnhofen},
  \bibinfo{person}{Hsiang-Ting Chen}, \bibinfo{person}{Patrick Baudisch},
  {et~al\mbox{.}}} \bibinfo{year}{2015}\natexlab{}.
\newblock \showarticletitle{Low-fidelity fabrication: Speeding up design
  iteration of 3D objects}. In \bibinfo{booktitle}{\emph{Proceedings of the
  33rd Annual ACM Conference Extended Abstracts on Human Factors in Computing
  Systems}}. \bibinfo{pages}{327--330}.
\newblock


\bibitem[\protect\citeauthoryear{Mueller, Im, Gurevich, Teibrich, Pfisterer,
  Guimbreti{\`e}re, and Baudisch}{Mueller et~al\mbox{.}}{2014}]%
        {mueller2014wireprint}
\bibfield{author}{\bibinfo{person}{Stefanie Mueller}, \bibinfo{person}{Sangha
  Im}, \bibinfo{person}{Serafima Gurevich}, \bibinfo{person}{Alexander
  Teibrich}, \bibinfo{person}{Lisa Pfisterer}, \bibinfo{person}{Fran{\c{c}}ois
  Guimbreti{\`e}re}, {and} \bibinfo{person}{Patrick Baudisch}.}
  \bibinfo{year}{2014}\natexlab{}.
\newblock \showarticletitle{WirePrint: 3D printed previews for fast
  prototyping}. In \bibinfo{booktitle}{\emph{Proceedings of the 27th annual ACM
  symposium on User interface software and technology}}.
  \bibinfo{pages}{273--280}.
\newblock


\bibitem[\protect\citeauthoryear{Multi3D}{Multi3D}{2021}]%
        {multi3d}
\bibfield{author}{\bibinfo{person}{Multi3D}.} \bibinfo{year}{2021}\natexlab{}.
\newblock \bibinfo{title}{Electrifi Conductive Filament}.
\newblock
\newblock
\urldef\tempurl%
\url{https://www.multi3dllc.com/product/electrifi/}
\showURL{%
\tempurl}


\bibitem[\protect\citeauthoryear{Norouzi, Kinnula, and Iivari}{Norouzi
  et~al\mbox{.}}{2021}]%
        {norouzi2021making}
\bibfield{author}{\bibinfo{person}{Behnaz Norouzi}, \bibinfo{person}{Marianne
  Kinnula}, {and} \bibinfo{person}{Netta Iivari}.}
  \bibinfo{year}{2021}\natexlab{}.
\newblock \showarticletitle{Making Sense of 3D Modelling and 3D Printing
  Activities of Young People: A Nexus Analytic Inquiry}. In
  \bibinfo{booktitle}{\emph{Proceedings of the 2021 CHI Conference on Human
  Factors in Computing Systems}}. \bibinfo{pages}{1--16}.
\newblock


\bibitem[\protect\citeauthoryear{Oliphant}{Oliphant}{2006}]%
        {oliphant2006guide}
\bibfield{author}{\bibinfo{person}{Travis~E Oliphant}.}
  \bibinfo{year}{2006}\natexlab{}.
\newblock \bibinfo{booktitle}{\emph{A guide to NumPy}}.
  Vol.~\bibinfo{volume}{1}.
\newblock \bibinfo{publisher}{Trelgol Publishing USA}.
\newblock


\bibitem[\protect\citeauthoryear{Oster}{Oster}{2011}]%
        {oster2011visicut}
\bibfield{author}{\bibinfo{person}{Thomas Oster}.}
  \bibinfo{year}{2011}\natexlab{}.
\newblock \showarticletitle{Visicut: An application genre for lasercutting in
  personal fabrication}.
\newblock \bibinfo{journal}{\emph{RWTH Aachen Univ}} (\bibinfo{year}{2011}).
\newblock


\bibitem[\protect\citeauthoryear{Pr{\'e}vost, Whiting, Lefebvre, and
  Sorkine-Hornung}{Pr{\'e}vost et~al\mbox{.}}{2013}]%
        {prevost2013make}
\bibfield{author}{\bibinfo{person}{Romain Pr{\'e}vost}, \bibinfo{person}{Emily
  Whiting}, \bibinfo{person}{Sylvain Lefebvre}, {and} \bibinfo{person}{Olga
  Sorkine-Hornung}.} \bibinfo{year}{2013}\natexlab{}.
\newblock \showarticletitle{Make it stand: balancing shapes for 3D
  fabrication}.
\newblock \bibinfo{journal}{\emph{ACM Transactions on Graphics (TOG)}}
  \bibinfo{volume}{32}, \bibinfo{number}{4} (\bibinfo{year}{2013}),
  \bibinfo{pages}{1--10}.
\newblock


\bibitem[\protect\citeauthoryear{Ramakers, Anderson, Grossman, and
  Fitzmaurice}{Ramakers et~al\mbox{.}}{2016}]%
        {ramakers2016retrofab}
\bibfield{author}{\bibinfo{person}{Raf Ramakers}, \bibinfo{person}{Fraser
  Anderson}, \bibinfo{person}{Tovi Grossman}, {and} \bibinfo{person}{George
  Fitzmaurice}.} \bibinfo{year}{2016}\natexlab{}.
\newblock \showarticletitle{Retrofab: A design tool for retrofitting physical
  interfaces using actuators, sensors and 3d printing}. In
  \bibinfo{booktitle}{\emph{Proceedings of the 2016 CHI Conference on Human
  Factors in Computing Systems}}. \bibinfo{pages}{409--419}.
\newblock


\bibitem[\protect\citeauthoryear{Roumen, M{\"u}ller, and Baudisch}{Roumen
  et~al\mbox{.}}{2018}]%
        {roumen2018grafter}
\bibfield{author}{\bibinfo{person}{Thijs~Jan Roumen}, \bibinfo{person}{Willi
  M{\"u}ller}, {and} \bibinfo{person}{Patrick Baudisch}.}
  \bibinfo{year}{2018}\natexlab{}.
\newblock \showarticletitle{Grafter: Remixing 3D-printed machines}. In
  \bibinfo{booktitle}{\emph{Proceedings of the 2018 CHI Conference on Human
  Factors in Computing Systems}}. \bibinfo{pages}{1--12}.
\newblock


\bibitem[\protect\citeauthoryear{Saakes, Cambazard, Mitani, and
  Igarashi}{Saakes et~al\mbox{.}}{2013}]%
        {saakes2013paccam}
\bibfield{author}{\bibinfo{person}{Daniel Saakes}, \bibinfo{person}{Thomas
  Cambazard}, \bibinfo{person}{Jun Mitani}, {and} \bibinfo{person}{Takeo
  Igarashi}.} \bibinfo{year}{2013}\natexlab{}.
\newblock \showarticletitle{PacCAM: material capture and interactive 2D packing
  for efficient material usage on CNC cutting machines}. In
  \bibinfo{booktitle}{\emph{Proceedings of the 26th annual ACM symposium on
  User interface software and technology}}. \bibinfo{pages}{441--446}.
\newblock


\bibitem[\protect\citeauthoryear{Savage, Head, Hartmann, Goldman, Mysore, and
  Li}{Savage et~al\mbox{.}}{2015}]%
        {savage2015lamello}
\bibfield{author}{\bibinfo{person}{Valkyrie Savage}, \bibinfo{person}{Andrew
  Head}, \bibinfo{person}{Bj{\"o}rn Hartmann}, \bibinfo{person}{Dan~B Goldman},
  \bibinfo{person}{Gautham Mysore}, {and} \bibinfo{person}{Wilmot Li}.}
  \bibinfo{year}{2015}\natexlab{}.
\newblock \showarticletitle{Lamello: Passive acoustic sensing for tangible
  input components}. In \bibinfo{booktitle}{\emph{Proceedings of the 33rd
  Annual ACM Conference on Human Factors in Computing Systems}}.
  \bibinfo{pages}{1277--1280}.
\newblock


\bibitem[\protect\citeauthoryear{Schmidt and Singh}{Schmidt and Singh}{2010}]%
        {schmidt2010meshmixer}
\bibfield{author}{\bibinfo{person}{Ryan Schmidt} {and} \bibinfo{person}{Karan
  Singh}.} \bibinfo{year}{2010}\natexlab{}.
\newblock \showarticletitle{Meshmixer: an interface for rapid mesh
  composition}.
\newblock In \bibinfo{booktitle}{\emph{ACM SIGGRAPH 2010 Talks}}.
  \bibinfo{pages}{1--1}.
\newblock


\bibitem[\protect\citeauthoryear{Schmidt and Umetani}{Schmidt and
  Umetani}{2014}]%
        {schmidt2014branching}
\bibfield{author}{\bibinfo{person}{Ryan Schmidt} {and}
  \bibinfo{person}{Nobuyuki Umetani}.} \bibinfo{year}{2014}\natexlab{}.
\newblock \showarticletitle{Branching support structures for 3D printing}.
\newblock In \bibinfo{booktitle}{\emph{ACM SIGGRAPH 2014 Studio}}.
  \bibinfo{pages}{1--1}.
\newblock


\bibitem[\protect\citeauthoryear{Sethapakdi, Anderson, Sy, and
  Mueller}{Sethapakdi et~al\mbox{.}}{2021}]%
        {fabricaide2021}
\bibfield{author}{\bibinfo{person}{Ticha Sethapakdi}, \bibinfo{person}{Daniel
  Anderson}, \bibinfo{person}{Adrian Reginald~Chua Sy}, {and}
  \bibinfo{person}{Stefanie Mueller}.} \bibinfo{year}{2021}\natexlab{}.
\newblock \bibinfo{booktitle}{\emph{Fabricaide: Fabrication-Aware Design for 2D
  Cutting Machines}}.
\newblock \bibinfo{publisher}{Association for Computing Machinery},
  \bibinfo{address}{New York, NY, USA}.
\newblock
\showISBNx{9781450380966}
\urldef\tempurl%
\url{https://doi.org/10.1145/3411764.3445345}
\showURL{%
\tempurl}


\bibitem[\protect\citeauthoryear{Slic3r}{Slic3r}{2011}]%
        {slic3r}
\bibfield{author}{\bibinfo{person}{Slic3r}.} \bibinfo{year}{2011}\natexlab{}.
\newblock \bibinfo{title}{Open source 3D printing toolbox}.
\newblock
\newblock
\urldef\tempurl%
\url{https://slic3r.org/}
\showURL{%
\tempurl}


\bibitem[\protect\citeauthoryear{Ultimaker}{Ultimaker}{2020}]%
        {curaUltimaker}
\bibfield{author}{\bibinfo{person}{Ultimaker}.}
  \bibinfo{year}{2020}\natexlab{}.
\newblock \bibinfo{title}{Ultimaker Cura}.
\newblock
\newblock
\urldef\tempurl%
\url{https://ultimaker.com/software/ultimaker-cura}
\showURL{%
\tempurl}


\bibitem[\protect\citeauthoryear{Umetani and Schmidt}{Umetani and
  Schmidt}{2013}]%
        {umetani2013cross}
\bibfield{author}{\bibinfo{person}{Nobuyuki Umetani} {and}
  \bibinfo{person}{Ryan Schmidt}.} \bibinfo{year}{2013}\natexlab{}.
\newblock \showarticletitle{Cross-sectional structural analysis for 3D printing
  optimization.}. In \bibinfo{booktitle}{\emph{SIGGRAPH Asia Technical
  Briefs}}. Citeseer, \bibinfo{pages}{5--1}.
\newblock


\bibitem[\protect\citeauthoryear{Vanek, Galicia, and Benes}{Vanek
  et~al\mbox{.}}{2014}]%
        {vanek2014clever}
\bibfield{author}{\bibinfo{person}{Juraj Vanek}, \bibinfo{person}{Jorge
  A~Garcia Galicia}, {and} \bibinfo{person}{Bedrich Benes}.}
  \bibinfo{year}{2014}\natexlab{}.
\newblock \showarticletitle{Clever support: Efficient support structure
  generation for digital fabrication}. In \bibinfo{booktitle}{\emph{Computer
  graphics forum}}, Vol.~\bibinfo{volume}{33}. Wiley Online Library,
  \bibinfo{pages}{117--125}.
\newblock


\bibitem[\protect\citeauthoryear{Wall, Jacobson, Vogel, and Schneider}{Wall
  et~al\mbox{.}}{2021}]%
        {wall2021scrappy}
\bibfield{author}{\bibinfo{person}{Ludwig~Wilhelm Wall}, \bibinfo{person}{Alec
  Jacobson}, \bibinfo{person}{Daniel Vogel}, {and} \bibinfo{person}{Oliver
  Schneider}.} \bibinfo{year}{2021}\natexlab{}.
\newblock \showarticletitle{Scrappy: Using Scrap Material as Infill to Make
  Fabrication More Sustainable}. In \bibinfo{booktitle}{\emph{Proceedings of
  the 2021 CHI Conference on Human Factors in Computing Systems}}.
  \bibinfo{pages}{1--12}.
\newblock


\bibitem[\protect\citeauthoryear{Wang, Chao, Tong, Yang, Tong, Li, Liu, and
  Liu}{Wang et~al\mbox{.}}{2015}]%
        {wang2015saliency}
\bibfield{author}{\bibinfo{person}{Weiming Wang}, \bibinfo{person}{Haiyuan
  Chao}, \bibinfo{person}{Jing Tong}, \bibinfo{person}{Zhouwang Yang},
  \bibinfo{person}{Xin Tong}, \bibinfo{person}{Hang Li},
  \bibinfo{person}{Xiuping Liu}, {and} \bibinfo{person}{Ligang Liu}.}
  \bibinfo{year}{2015}\natexlab{}.
\newblock \showarticletitle{Saliency-preserving slicing optimization for
  effective 3D printing}. In \bibinfo{booktitle}{\emph{Computer Graphics
  Forum}}, Vol.~\bibinfo{volume}{34}. Wiley Online Library,
  \bibinfo{pages}{148--160}.
\newblock


\bibitem[\protect\citeauthoryear{Weichel, Alexander, Karnik, and
  Gellersen}{Weichel et~al\mbox{.}}{2015}]%
        {weichel2015spata}
\bibfield{author}{\bibinfo{person}{Christian Weichel}, \bibinfo{person}{Jason
  Alexander}, \bibinfo{person}{Abhijit Karnik}, {and} \bibinfo{person}{Hans
  Gellersen}.} \bibinfo{year}{2015}\natexlab{}.
\newblock \showarticletitle{SPATA: Spatio-tangible tools for fabrication-aware
  design}. In \bibinfo{booktitle}{\emph{Proceedings of the Ninth International
  Conference on Tangible, Embedded, and Embodied Interaction}}.
  \bibinfo{pages}{189--196}.
\newblock


\bibitem[\protect\citeauthoryear{Zhang, Le, Panotopoulou, Whiting, and
  Wang}{Zhang et~al\mbox{.}}{2015}]%
        {zhang2015perceptual}
\bibfield{author}{\bibinfo{person}{Xiaoting Zhang}, \bibinfo{person}{Xinyi Le},
  \bibinfo{person}{Athina Panotopoulou}, \bibinfo{person}{Emily Whiting}, {and}
  \bibinfo{person}{Charlie~CL Wang}.} \bibinfo{year}{2015}\natexlab{}.
\newblock \showarticletitle{Perceptual models of preference in 3D printing
  direction}.
\newblock \bibinfo{journal}{\emph{ACM Transactions on Graphics (TOG)}}
  \bibinfo{volume}{34}, \bibinfo{number}{6} (\bibinfo{year}{2015}),
  \bibinfo{pages}{1--12}.
\newblock


\bibitem[\protect\citeauthoryear{Zhao, Gu, Huang, Garcia, Chen, Tu, Benes,
  Zhang, Cohen-Or, and Chen}{Zhao et~al\mbox{.}}{2016}]%
        {zhao2016connected}
\bibfield{author}{\bibinfo{person}{Haisen Zhao}, \bibinfo{person}{Fanglin Gu},
  \bibinfo{person}{Qi-Xing Huang}, \bibinfo{person}{Jorge Garcia},
  \bibinfo{person}{Yong Chen}, \bibinfo{person}{Changhe Tu},
  \bibinfo{person}{Bedrich Benes}, \bibinfo{person}{Hao Zhang},
  \bibinfo{person}{Daniel Cohen-Or}, {and} \bibinfo{person}{Baoquan Chen}.}
  \bibinfo{year}{2016}\natexlab{}.
\newblock \showarticletitle{Connected fermat spirals for layered fabrication}.
\newblock \bibinfo{journal}{\emph{ACM Transactions on Graphics (TOG)}}
  \bibinfo{volume}{35}, \bibinfo{number}{4} (\bibinfo{year}{2016}),
  \bibinfo{pages}{1--10}.
\newblock


\bibitem[\protect\citeauthoryear{Zhou and Jacobson}{Zhou and Jacobson}{2016}]%
        {zhou2016thingi10k}
\bibfield{author}{\bibinfo{person}{Qingnan Zhou} {and} \bibinfo{person}{Alec
  Jacobson}.} \bibinfo{year}{2016}\natexlab{}.
\newblock \showarticletitle{Thingi10k: A dataset of 10,000 3d-printing models}.
\newblock \bibinfo{journal}{\emph{arXiv preprint arXiv:1605.04797}}
  (\bibinfo{year}{2016}).
\newblock


\end{thebibliography}

\appendix

\end{document}